\documentclass[12p, times]{FAB-2018}

\usepackage{bibentry}
\usepackage{float}

\usepackage{graphicx,url}
\usepackage{hyperref}

\usepackage[english]{babel}   
\usepackage[utf8]{inputenc}
\nobibliography*

\usepackage{listings}
\usepackage{color}
 
\lstdefinestyle{numbexrs}{numberstyle=\tiny}
\title{A Case Study for Grain Quality Assurance Tracking based on a Blockchain Business Network}

\author{Percival Lucena \\ IBM Research  
   \and Alecio P. D. Binotto \\ IBM Research
   \and Fernanda da Silva Momo  \\  UFRGS
    \and Henry Kim  \\  York University
 }

\begin{document}

\maketitle

\begin{abstract}
One of the key processes in Agriculture is quality measurement throughout the transportation of grains along its complex supply chain. This procedure is suitable for failures, such as delays to final destinations, poor monitoring, and frauds. To address the grain quality measurement challenge through the  transportation chain, novel technologies, such as Distributed Ledger and Blockchain, can bring more efficiency and resilience to the process. Particularly, Blockchain is a new type of distributed database in which transactions are securely appended using cryptography and hashed pointers. Those transactions can be generated and ruled by special network-embedded software -- known as smart contracts -- that may be public to all nodes of the network or may be private to a specific set of peer nodes. This paper analyses the implementation of Blockchain technology targeting grain quality assurance tracking in a real scenario. Preliminary results support a potential demand for a Blockchain-based certification that would lead to an added valuation of around 15\%  for GM-free soy in the scope of a Grain Exporter Business Network in Brazil.

\end{abstract}
     
\section{Introduction}

According to a WorldBank Report \cite{worldbank}, in 2015 Brazil exported 195 billion dollars, making it the 21st largest exporter in the world. Exports are led by soybeans which represent 11\% of the total exports of Brazil. Other grains such as coffee 3\% and corn 2.6\% also represent a significant percentage of Brazilian exports. 

The United States Department of Agriculture annual production, supply and distribution report states that  Brazil is the second largest soybean producer in the world, behind only the United States\cite{USDA:2017:Online}. The sum of exports from January to August 2017 is over 57 million tonnes \cite{Conab:2017:Online}. In comparison with 2016, in this same period, there was an increase of more than 8.7 million tons exported.

The Brazilian cereal grains production for the 2016/17 harvest increased by  28\% reaching 238.8 million tons as reported by the 12th Grain Survey conducted by Conab (Brazilian National Supply Company) \cite{Conab:2017:Online}.  A FIESP survey (Federation of Industries of the State of São Paulo) shows that the planted area dedicated to grains in Brazil is estimated at 60.9 million hectares which represents the largest area registered in the  historical series \cite{FIESP:2017:Online}.

The increasing productivity of  grains  in Brazil is related to a greater use of technology in the field, present not only in machines and implements but also in seeds, cultivation techniques and also the use of irrigation. Regarding the production of maize, soybean, and wheat, there is a variation of productivity in relation to the 15/16 harvest, respectively, 32.9\%, 17.2\% and -14.8\% \cite{Conab:2017:Online}.  

Forasmuch as technology has helped Brazilian cereals grain production,  agriculturalists still face  logistics and warehouse challenges to move the production from the farm to port terminals and processing industries. More than sixty percent of the Brazilian soybean production is transported by truck from the production areas to the port terminals \cite{danao}.   

The existing transportation and warehouse storage processes often affect grain quality causing grain damage, moisture, and contamination \cite{caixeta2008}. Grain quality control information is often kept in spreadsheets and spread among diverse ledgers which often provides inaccurate information causing financial losses on negotiation between producers and traders.

This case study aims to describe and highlight the gains obtained with the implementation of a Blockchain Business Network for Brazilian Agriculture exports; the lessons learned from the implementation of this project; the challenges in the development of Blockchain platform and future opportunities of using Blockchain in other contexts.

This paper presents the GEBN Blockchain Business Network, an  enterprise consortium  that aggregates data from certified quality assurance processes and provides information for diverse business partners of The Brazilian Grain Exporters Business Network. This platform  helps producers track grains stored in warehouses  optimizing  trading with global exporters. The remainder of this paper will explain the solution implemented and it is structured as follows: Section 2 introduces Blockchain Business Networks.  Section 3 describes the Brazilian Grain Exporters Business Network.  Section 4 presents the case study context. Section 5 presents a Proof of Concept developed, and Section 6 presents results and conclusions. 
\\

\section{Blockchain Technology Applications on Agriculture}
\label{sec:background}

 Blockchain can be seen as a ``disruptive innovation with a wide range of applications, potentially capable of redesigning our interactions in business, politics, and society in general'' \cite{atzori2015blockchain} . Focusing on the business field, Cohen, Amorós, and Lundy (2017) \cite{cohen2017generative} point out that the use of Blockchain in solving different business problems, from different segments, allows for modifications in existing business models and even the creation of new business models. Therefore, this technology allows new opportunities for creating customer value in a business model suitable for its exploration \cite{cohen2017generative}.

Blockchain has originated as a shared database for recording the history of Bitcoin transactions \cite{nakamoto2008}. These transactions are grouped in blocks,  including  hashed pointers to previous blocks, that provide the accepted history of transactions  since the inception of the Blockchain. This architecture has been implemented and extended by several general purpose ledgers such as Hyperledger Fabric \cite{cachin2016architecture}, R3 Corda \cite{brown2016introducing} and Ethereum \cite{wood2014ethereum}.

The Blockchain can be regarded as a complex, network-based software connector, which provides communication, coordination (through transactions, smart contracts, and validation oracles) and facilitation services. Every node in the blockchain network has two layers, namely, application layer and blockchain layer. Part of the application is implemented inside the blockchain connector in terms of smart contracts \cite{xu2016blockchain}.

Szabo \cite{szabo1997formalizing} has coined the ``smart contract'' term as software representation for many kinds of contractual clauses in a way to make a breach of contract expensive for the breacher.  Smart contracts could be used to represent liens, bonding, delineation of property rights, and other paper-based contracts.  Smart contracts may be operated by a consortium comprising of parties in a multilateral contract \cite{kim2017}.


\subsection{Blockchain Business Networks}

Advanced globalization has created several complex supply chain scenarios where different companies cooperate to form a ``quasi-organization'' \cite{HAKANSSON2002133}. A Business Network describes the structures and processes that exist in the exchange of assets among participants in economic networks.   

In the business network, every company “connects different people, various activities and miscellaneous resources with varying degrees of mutual fit” and “operates within a texture of interdependencies that affects its development” \cite{snehota1995developing}. Technology, knowledge, social relations, administrative routines, and systems are some resources that are encountered in business relationships. 

Therefore, participants, assets, registries, and transactions are shown as fundamental elements of a business network. Participants are the actors in the business network and might be an individual or an organization. Assets are created by participants and subsequently exchanged between them through transactions. Assets can have a rich lifecycle, as defined by the transaction in which they are involved. As assets move through their lifecycle and through different registries they can be in more than one registry at the same time.

\begin{figure}[h!]
\centering
\includegraphics[scale=0.5]{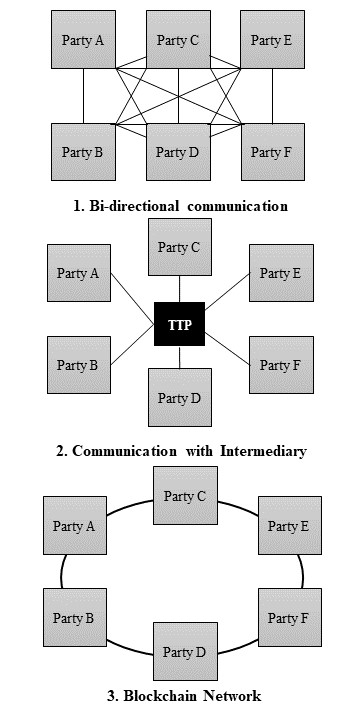}
\caption{External business collaborations implemented on Blockchain.}
\label{fig:cont-delivery}
\end{figure}

Complex business networks can be represented in different ways.  Previous to the advent of  Blockchchain, most multi-party business collaborations were implemented in the form of multiple binary relationships.  Alternatively, the business partners would rely on a Trusted Third Party (TTP) to facilitate interactions, since a TTP acts for building and repair the trust \cite{brodt2013repairing}. Another possibility is the use of technology as a way to reduce TTP's participation in the business network maintaining trusting.  Blockchain can represent complex business networks by nodes for each one of the different companies that need to cooperate and exchange information.

Contracts are usually created as a set of promises to formalize relationships in a Business Network. Whenever a trusted intermediary is removed, the organizations involved in  interdependent processes  must find alternative means to provide a division of labor as effectively as the displaced contractor.  Smart contracts can  provide such division of labor and near decomposability \cite{kim2017}.

Blockchain provides a new approach to Business Networks. According to several authors \cite{Valdes&Furlonger2016, Tapscott&Tapscott&Kirkland2016, McKendrick2017, Tapscott&Tapscott2017}, this technology has a great potential of impact and revolution in the world economy from the generation of changes in organizations and in the way business is done. Most of the blockchain networks assume organizations running the peers have no trust relationship established between them. Encryption, consensus, and other algorithms of blockchain guarantee trusted outcomes in this context \cite{Hull2016}. 

In this regard, Figure 1 presents the relationship configurations mentioned before between the parties involved in a transaction of a complex business network. We highlight from these configurations that the use of Blockchain allows a better flow between the members of the business network and remove the role of some intermediaries in some business processes \cite{nakamoto2008}.

\section{The Brazilian Grain Exporters Business Network (GEBN)}

The Grain Exporters Business Network (GEBN) in Brazil is composed of a diverse set of players including grain producers, rural credit cooperatives, warehouse companies, tradings exporters, agrochemical companies, freight forwarders, and ports authorities. There are different types of transactions and contracts among business partners for financing, sales, transportation, warehousing, among others.  

\begin{figure}[H]
\centering
\includegraphics[scale=0.45]{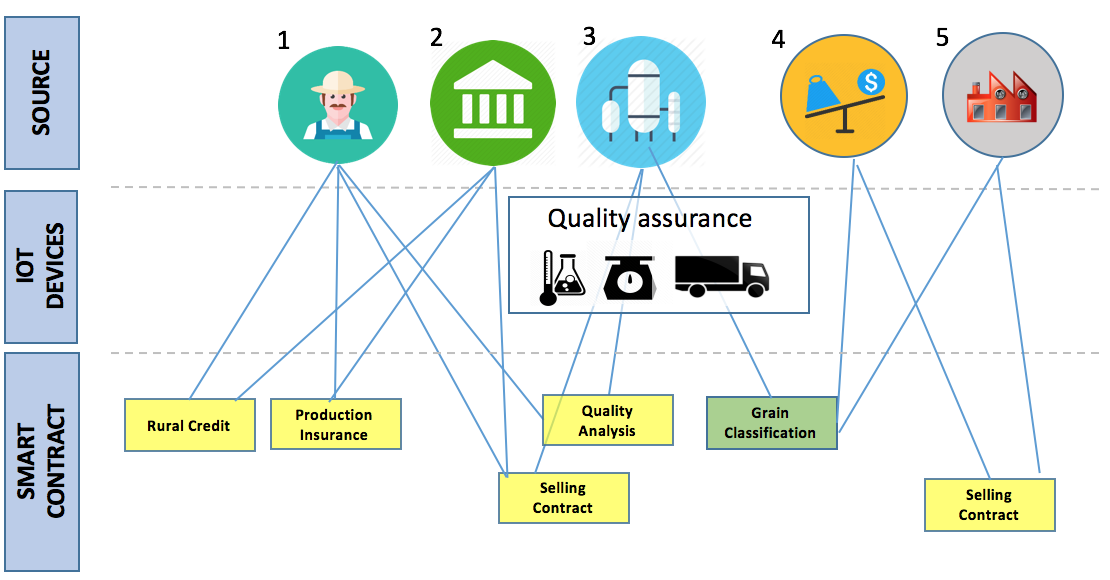}
\caption{Grain Exporters Business Network}\label{fig:blocks}
\end{figure}

 Figure 2 illustrates part of the Grain Exporters Business Network  consortium. Actor 1 represents a grain producer who is usually concerned that the grains ingest and classification  are properly conduced, so he can  be fairly paid.  The ingest receipt is also important for the Producer  in order to obtain credit on Rural Credit Banks and to contract production insurance as well. Actor 2  represents a Rural Credit Bank agent who depends on accurate data from producers  to reduce credit giving risks and offer  lower interest credit rates for credit operations.  Actor 3  represents a private warehouse agent who depends on accurate grain  classification data in order to wholesale grains. Actor 4 represents a trading company agent who is concerned in buying large amount of grains with the right quality from several warehouses in order to fulfill an export request. Actor 5 represents a food processing company who is concerned in buying special selected grains that have specific characteristics such as high protein, high carbohydrates and low moisture levels.

The different players from the grain supply chain were not able to trust a centralized system due to the different business goals  of the business actors. The Brazilian Government  National Supply Company (CONAB) GeoSafras Report \cite{figueiredo2005projeto} offers yearly total estimates of grains production. Unfortunately, real time information about grain products is segregated to local actors in the supply chain.

Blockchain smart contracts provides a  fair trading system of exchange that honours producers, communities, wholesalers, traders and the environment.  Blockchain immutable ledger  provides a compliance mechanism in order to avoid frauds in the grain classification process.  As described in Figure 2, GEBN platform was created so that laboratory quality assurance test devices can connect directly to the Blockchain so the process information cannot be changed. This provides extra trust for the GEBN business partners who depend on accurate grain information. Blockchain can increase its likelihood of export to international markets since compliance with international standards becomes a transparent and undisputed matter \cite{lin2017blockchain}. 
 
The  designed solution plans to be extended to cover most common agriculture processes allowing Agribusiness partners to collaborate seamlessly, reducing communication time, helping to track product provenance, reducing costs, and proving trust among GEBN partners.

\section{Case Study context}

The grain mass stored in the silos deteriorates in relation to the interaction between physical variables (temperature, humidity, warehouse structure, meteorological variables), chemical variables (oxygen availability in intergranular air) and biological variables of internal sources (longevity, respiration, post-harvest maturity and germination) and biological variables from external sources (fungi, yeasts, bacteria, insects, mites, rodents and birds). The degree of deterioration depends on the rate of increase of these variables which, in turn, are mainly affected by the interaction of temperature and humidity and secondarily by their interrelation with the grain, between them, and with the structure of the silo \cite{sinhagrain}.

The search for the quality of grains and by-products is a priority for producers, processors, and for distributors of these products. Quality assurance processes are used on GEBN to provide grain classification according to international standards. Due to deterioration factors, it is important to have accurate information of the grains when they arrive and leave the warehouse silos. The intrinsic and extrinsic analysis processes are executed on both incoming and outgoing trucks, so grains quality assurance compliance process provides decision information for both grains buyers and sellers. 

In a standard quality assurance process, an automatic crane located in the warehouse entrance,  removes samples from a grain loaded truck. Those samples are sent by pipes to a packer in the laboratory. A laboratory technician collects the samples and submits them to intrinsic and extrinsic analysis tests.

\subsection {Intrinsic Analysis}

This process analyses soybeans and corn samples in order to detect if the grains are genetically modified (GMO) and if mycotoxins levels are at acceptable levels. Sample grains are ground so that 60-70\% of the sample must pass through a 20-mesh sieve. Grains are then  mixed with water in a (1:5) proportion. A 12 ml sample is removed from the solution using a pipette and dispensed into a reaction cup where a reaction strip is placed for 5 minutes.  Inside the reaction tube, the sample travels by capillary action of an end to the other extreme of the tape. When passing through the membrane, the sample comes in contact with the antibodies that react with the target analyte \cite{envirologix}.

After the testing period, the strip is then immediately placed into Envirologix QuickScanner device which is connected to a computer that reads the information and stores it on GEBN Blockchain.  Each batch of test strips for GMOs or mycotoxins is tested and compared to known quantification patterns generating specific batch curves. The standard curve data is encoded in a 2-D strip itself. When the strip is read, Envirologix QuickScanner software measures the information of the standard curve present in the barcode and calculates the amount of analyte specific to each test tape.

\subsection {Extrinsic Analysis}

Brooker et al. \cite{brooker1992} considers several properties for extrinsic analysis of grains, such as: moisture content, specific mass, the percentage of broken grains, impurities, damages caused by drying temperature, susceptibility to breakage, grinding, the presence of insects and fungi, type of grain and year of production. GEBN Blockchain extrinsic analysis process determines some of those grains characteristics such as moisture levels, broken and damage levels. 

Gehaka Moisture Analyzers installed in the Quality assurance laboratory provide information directly to the GEBN Blockchain. The quality process that requires visual analysis is stored including information of the testing operator and exam date in order to avoid disputes.

\section{Experiment}

On our GEBN study, each business partner owns a node in the network with a full copy of transaction data shared among all the participants.  Three nodes were created on GEBN: one for the cooperative producers, another for the warehouse originator company and a third for a rural credit bank. 

GEBN was deployed on a Hyperledger Fabric  Blockchain Cloud instance. Figure 3 illustrates quality assurance transactions stored on GEBN Blockchain. Hyperledger 1.0 query mechanisms are used to trace the origin of an outgoing lot helping audit processes through the supply chain.  Hyperleger Fabric 1.0 blockchain also provides private communication channels, allowing computation to occur only among business partner nodes involved in a transaction.  Permissioned blockchains such as Hyperledger Fabric have a set of trusted parties to carry out verification, and additional verifiers can be added to the agreement of the current members of the consortium. \cite{peters2016understanding}.   Permissioned blockchains offer clear advantages in security and privacy while potentially reducing costs of compliance with regulations \cite{Yermack2017}.

\begin{figure}[H]
\centering
\includegraphics[scale=0.42]{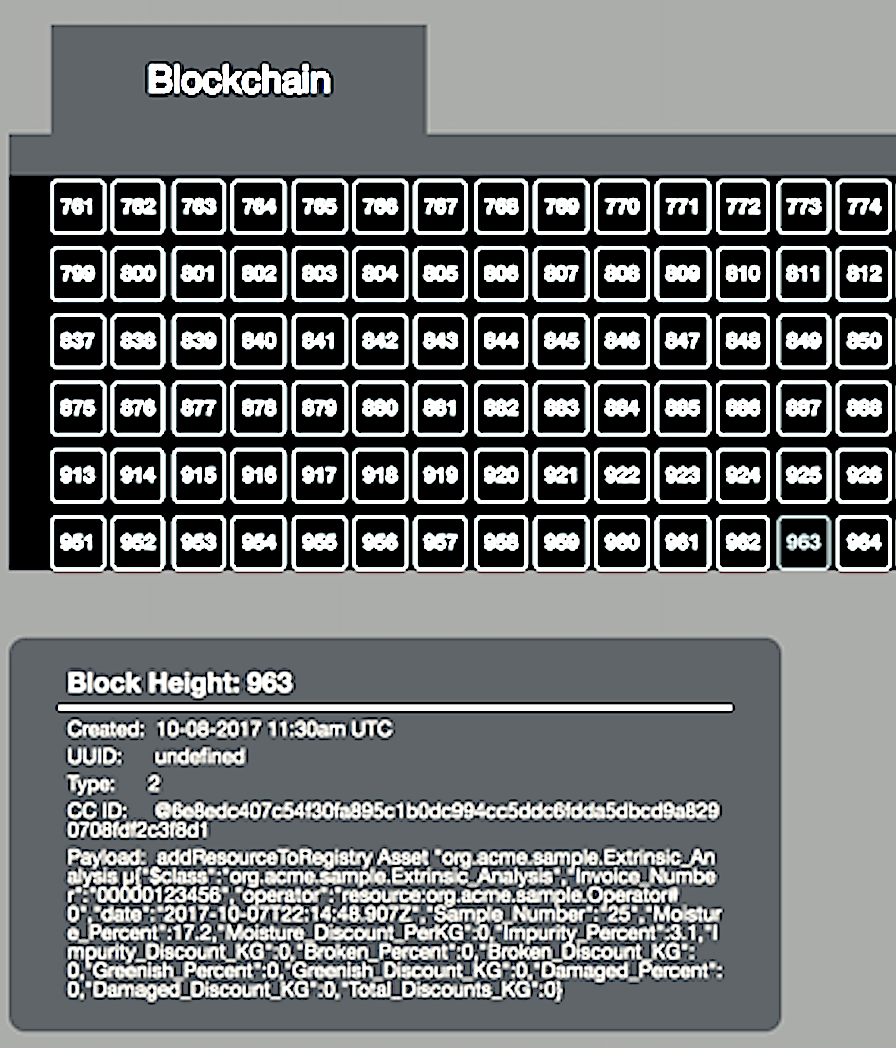}
\caption{Quality Assurance Transaction on GEBN Blockchain.}
\label{fig:blocks}
\end{figure}

Smart contracts were  created using the Hyperledger Composer Framework \cite{kakadiya2017block}  as described on Figure 4. Hyperledger Composer Framework includes a standalone Node.js process that exposes a business network as a LoopBack REST API  \cite{mardan2014sails}. The communication between the Grain Controller Desktop Application (GDPA) and the GEBN Blockchain server is protected by  PassportJS open source authentication middleware \cite{pereira2016authenticating} configured with passport-local strategy that implements an access control list.

\begin{figure}[H]
\centering
\includegraphics[scale=0.4]{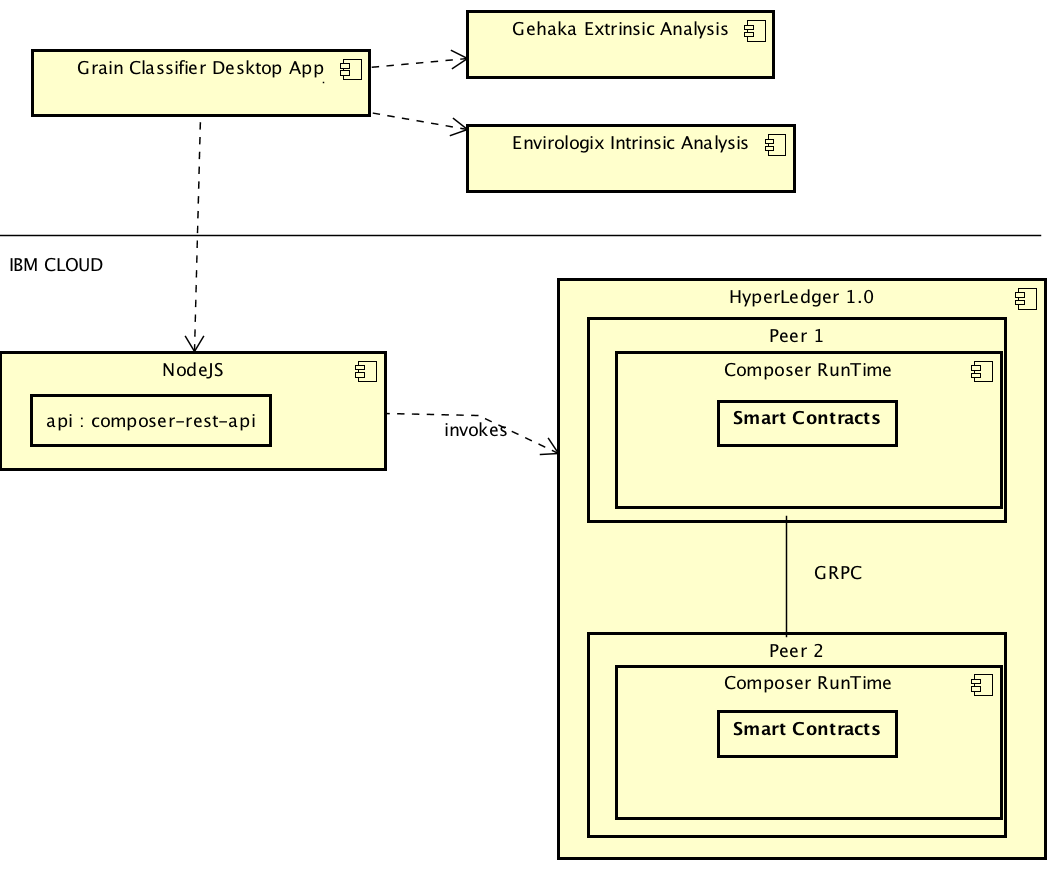}
\caption{High Level Architecture}
\label{fig:blocks}
\end{figure}

The Grain Controller Desktop Application  was installed at a single Warehouse Quality Control lab  in Ribeirão do Sul, Brazil. All the grains received at this facility pass through the quality control procedure before they are stored in silos. The quality control process allows the segregation of grains by levels of quality so prices are set according to the characteristics of the stored and marketed cargo.  Listing 1 shows a sample smart contract developed for calculating price discounts for soybeans.

\begin{figure}[H]
\centering
\includegraphics[scale=0.32]{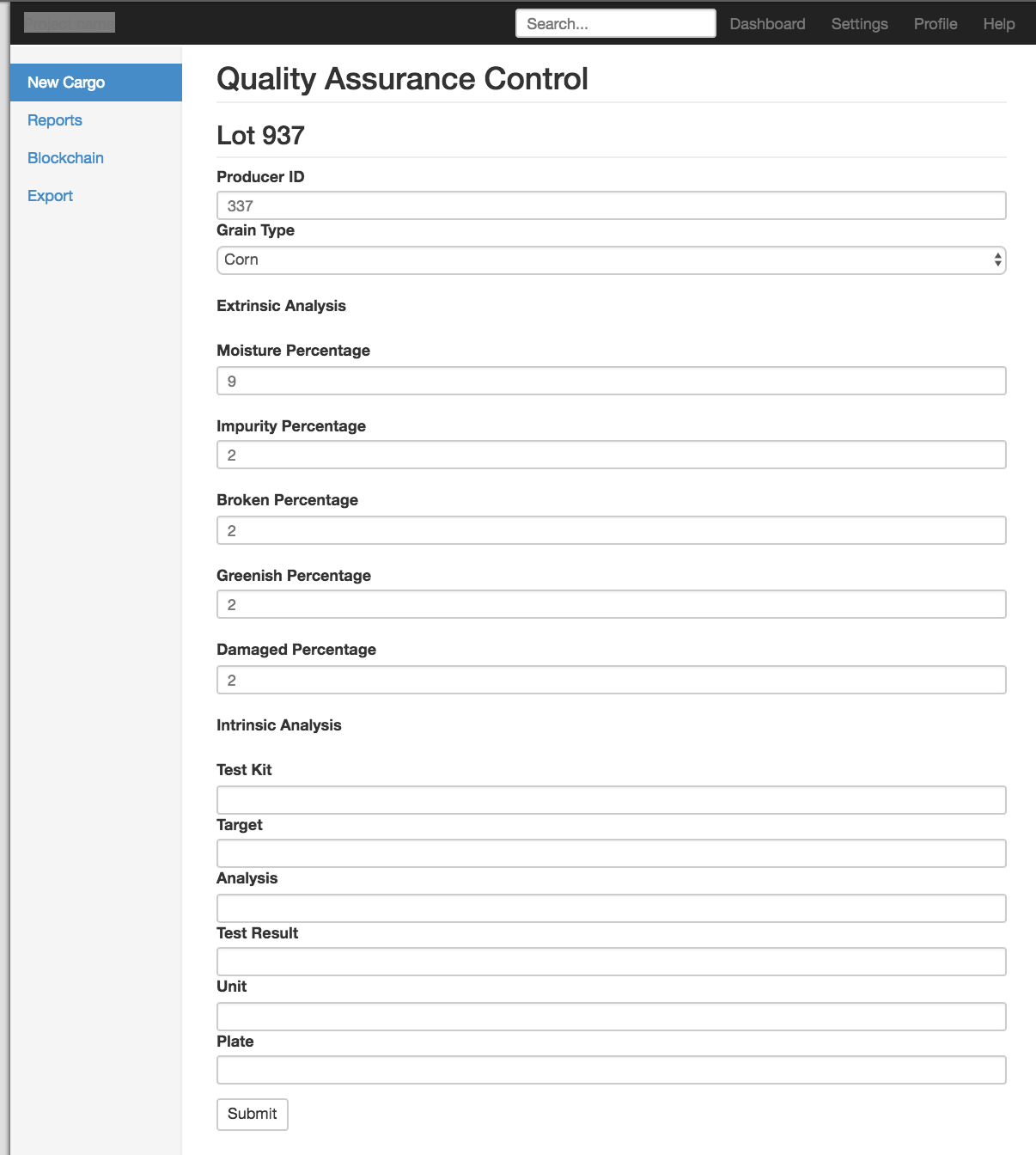}
\caption{Grain Controller Desktop Application}
\label{fig:frontend}
\end{figure}

The quality control procedure starts when the trucks arrive at the warehouse facility. The cargo electronic weighting information is stored on GEBN Blockchain. Afterwards, samples are taken from all incoming and outgoing trucks. Those samples are submitted to extrinsic and intrinsic analysis processes.

A certified quality assurance user authenticates in GDPA and  inputs information about the producer's cargo that arrives at the facility as described on Figure 4. GDPA reads quality data directly from Intrincic and Extrinsic analysis devices.  Producer's identity is also stored  on GEBN Blockchain. The analysis results determine the pre-cleaning and drying process for the cargo and the appropriate storage silo. When a grain cargo is sold, a similar process is executed on the outgoing truck.

\begin{lstlisting}[language=Java,caption=DiscountsTransaction]

01. asset Extrinsic_Analysis identified by Invoice_Number {
02.    o String Invoice_Number
03.    --> Operator operator
04.    o DateTime date
05.    o String Sample_Number
06.    o Double Moisture_Percent
07.    o Double Impurity_Percent
08.    o Double Broken_Percent
09.    o Double Greenish_Percent
10.    o Double Damaged_Percent
11.    o Double Total_Discounts_KG
12. }
13.
14. transaction DiscountsTransaction {
15.  --> Extrinsic_Analysis asset
16. }
17.
18. event DiscountsEvent {
19.  --> Extrinsic_Analysis asset
20. }
21.
22. function discountsTransaction(tx) {
23.    var d=0;
24.    if(tx.asset.Moisture_Percent>12)
25.    d+=(tx.asset.Moisture_Percent-12)*4;
26.    if(tx.asset.Impurity_Percent>3) 
27.    d+=(tx.asset.Impurity_Percent-3)*2.5;
28.    if(tx.asset.Broken_Percent>5)
29.    d+=(tx.asset.Moisture_Percent-5)*1;
30.    if(tx.asset.Damaged_Percent>3) 
31.    d+=(tx.asset.Impurity_Percent-3)*3.5;  
32.    tx.asset.Total_Discounts_KG = d;
33.
34. return 
35. getAssetRegistry('com.agritech.Extrinsic_Analysis')
36.        .then(function (assetRegistry) {
37.            return assetRegistry.update(tx.asset);
38.        })
39.        .then(function () {
40.         var event =getFactory()
41.         .newEvent('com.agritech', 'DiscountsEvent');
42.         event.asset = tx.asset;
43.         tx.asset.Total_Discounts_KG;
44.         emit(event);
45.   });
46. }

\end{lstlisting}

\section{Conclusions}

This study aims to describe and highlight the gains obtained with the implementation of the blockchain platform in the agricultural context; the lessons learnt from the implementation of blockchain in the agricultural context; the challenges in the development of blockchain platform and future opportunities of using blockchain in other contexts.

One of the main advantages of using Blockchain, in spite of other software development platforms, is that all the members of the GEBN can now share the same business rules and  transaction data in their nodes reducing disputes among business partners, information asymmetries and consequently improving governance.   

The transactions transparency provided by Blockchain requires that the companies involved in the supply chain 
to collaborate effectively defining common rules that can be expressed in smart contracts. The formation of consortia like GEBN is an interesting form of organization that allows members of a supply chain to vote on the rules and consensus principles for transactions settlements. 

We found a controversial use of Blockchain regarding its legal value.  In order to prevent disputes, our approach focused on signing all transactions with an identity of a recognized member of the consortia. Brazilian legislation implemented by Medida Provisoria 2.200-2/2001  recognizes digital signatures on documents to have legal value. Nonetheless, complex scenarios involving international trade and arbitration laws are yet to be proven.

Although our blockchain application focused on grains quality control we believe there is a huge opportunity for Blockchain applications on  Global Trade. Trading partners today rely on a complex and paper-heavy process to secure their transactions. Buyers, sellers,  banks, transporters, inspectors, regulators have their own forms and records to fill out in separate systems of records.  Capital is tied up as paper documents are sent back and forth, checked and rechecked, reviewed and reconciled. Delays caused by errors and manual processing can make it difficult for companies to access financing, which could cause business inefficiencies. Blockchain can provide a shared version of the truth, so trade partners can interact with greater trust. Third-party verification processes could be simplified by the use of smart contracts reducing the potential for errors or tampering. This can increase the efficiency with which companies access funding as well as save time and costs throughout the trade process.

Like all prior disruptive technologies, there will be beneficial and detrimental aspects of Blockchain technologies that will be tested as the first Blockchain Business networks start to operate.  As the technology matures, Blockchain Business networks should provide several new business models that can revolutionize several industries worldwide. 

\bibliographystyle{unsrt}
\bibliography{r2}

\end{document}